\documentclass{appolb}
\usepackage{graphicx}

\usepackage[utf8x]{inputenc}
\usepackage[OT4]{fontenc}
\usepackage[english]{babel}
\usepackage{graphicx}
\usepackage{hyperref}
\usepackage{xcolor}
\usepackage{pifont}
\usepackage{ulem}

\usepackage{amsmath}
\begin{document}
\title{News on strangeness production from NA61/SHINE
\thanks{Presented at Quark Matter 2022, 5 April 2022, Krak\'ow, Poland}%
}
\author{
Maciej Lewicki
\address{Institute of Nuclear Physics Polish Academy of Sciences}
\\[3mm]
%Marjan Ćirković
%\address{University of Belgrade}
%\\[3mm]
%Magdalena Kuich, Piotr Podlaski
%\address{University of Warsaw}
%\\[3mm]
%Szymon Puławski
%\address{University of Silesia}
%\\[3mm]
%Angelika Tefelska
%\address{Warsaw University of Technology}
for the NA61/SHINE Collaboration
}
\maketitle
\begin{abstract}
NA61/SHINE is a fixed target experiment at the CERN Super Proton Synchrotron. The main goals of the experiment are to discover the critical point of strongly interacting matter and to study the properties of the onset of deconfinement. In order to reach these goals, a study of hadron production properties is performed in nucleus-nucleus, proton-proton and proton-nucleus interactions as a function of collision energy and size of the colliding nuclei.
In this talk, the new results on identified charged kaon production in the intermediate size system ($^{40}$Ar+$^{45}$Sc and $^7$Be+$^9$Be) collisions at SPS beam momentum range (13\textit{A}-150\textit{A} GeV/\textit{c}) will be shown. Additionally, the new measurements of strange mesons ($K^0_S$), resonances ($K^{0*}$(892), $\Xi(1530)^0$ and $\bar{\Xi}(1530)^0$) and baryons ($\Xi^-$(1321), $\Xi^+$(1321))  produced in $p$+$p$ interactions are presented.
\end{abstract}

\section{NA61/SHINE facility}
NA61/SHINE is a fixed target spectrometer \cite{1} located in CERN’s North Area, utilizing the SPS proton, ion and hadron beams. Tracking capabilities are provided by four large volume Time Projection Chambers (TPC), two of which are located in magnetic fields. The Projectile Spectator Detector (PSD), a zero degree, modular calorimeter, is used to determine the centrality of the collisions.
The aim of the experiment is to explore the QCD phase diagram ($\mu_B, T$) by a two-dimensional scan in collision energy and system size. The yields of hadrons produced in the collisions are studied for indications of the onset of deconfinement \cite{2} and the critical point of the phase transition \cite{3}. This article discusses the new NA61/SHINE results on the strange hadron production in two main categories: charged kaon spectra in intermediate size systems of Ar+Sc and Be+Be \cite{4,5,bebe1,bebe2,arsc1}, and the high precision measurements of strange mesons, baryons and resonances from $p$+$p$ interactions \cite{kstar1,kstar2,marjan,xi}. The data on strange hadron production properties are a rich source of knowledge on the dynamics of high energy collisions, particularly in the region of deconfinement phase transition.

\section{Studies of the onset of deconfinement by NA61/SHINE}

\begin{figure}
{\centering
	\includegraphics[width=.3\linewidth]{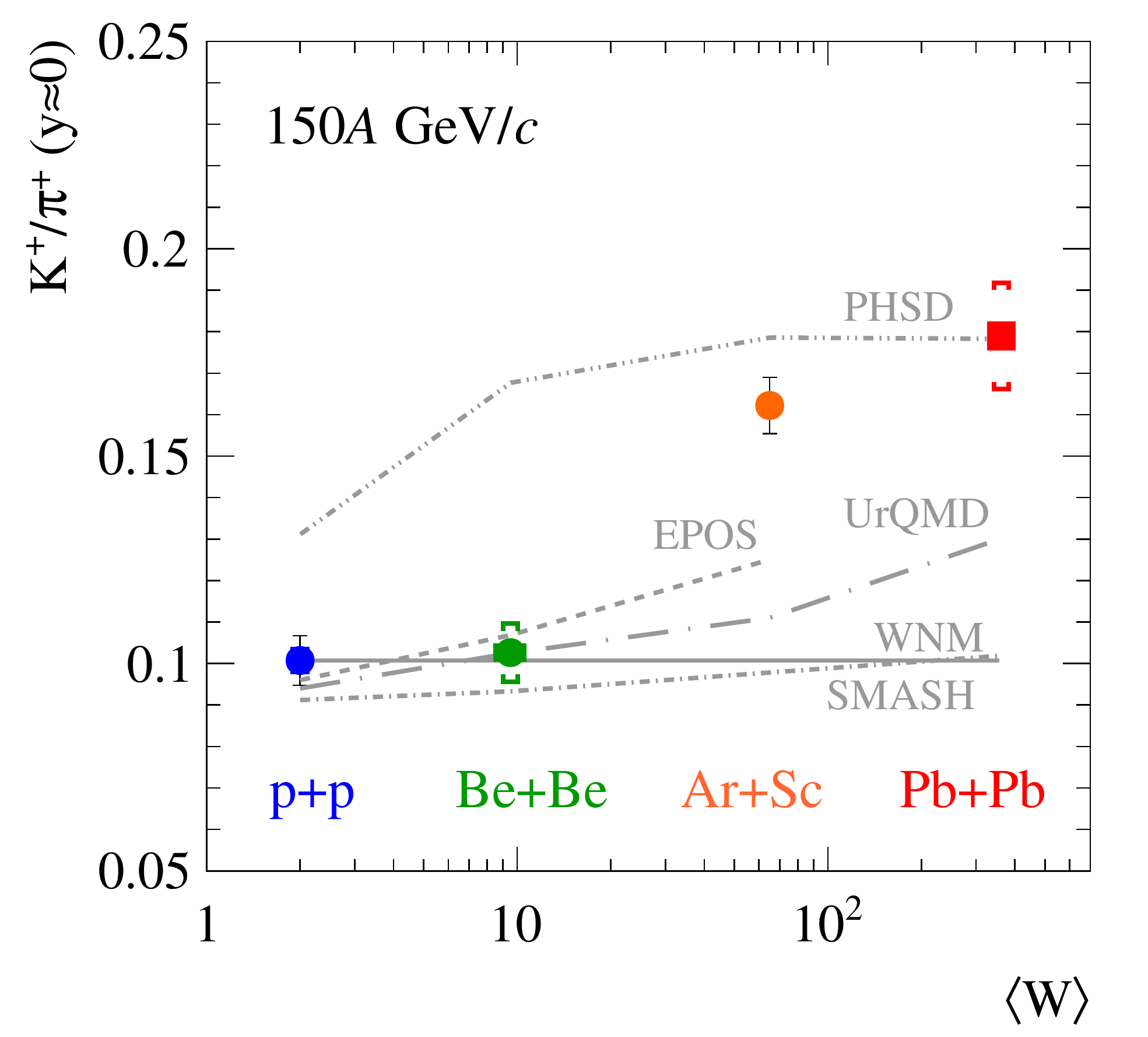}~
	\includegraphics[width=.3\linewidth]{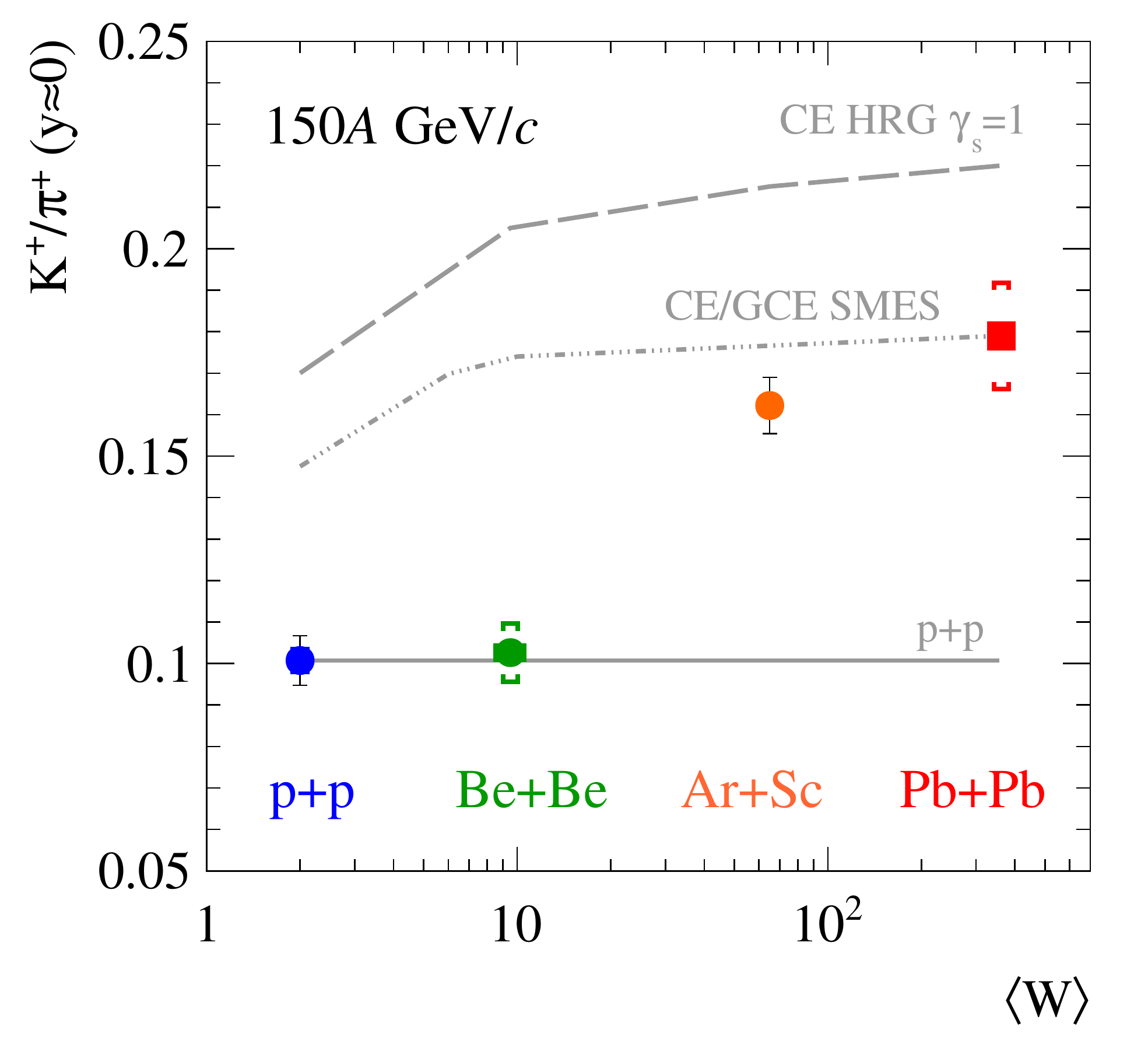}~
	\includegraphics[width=.3\linewidth]{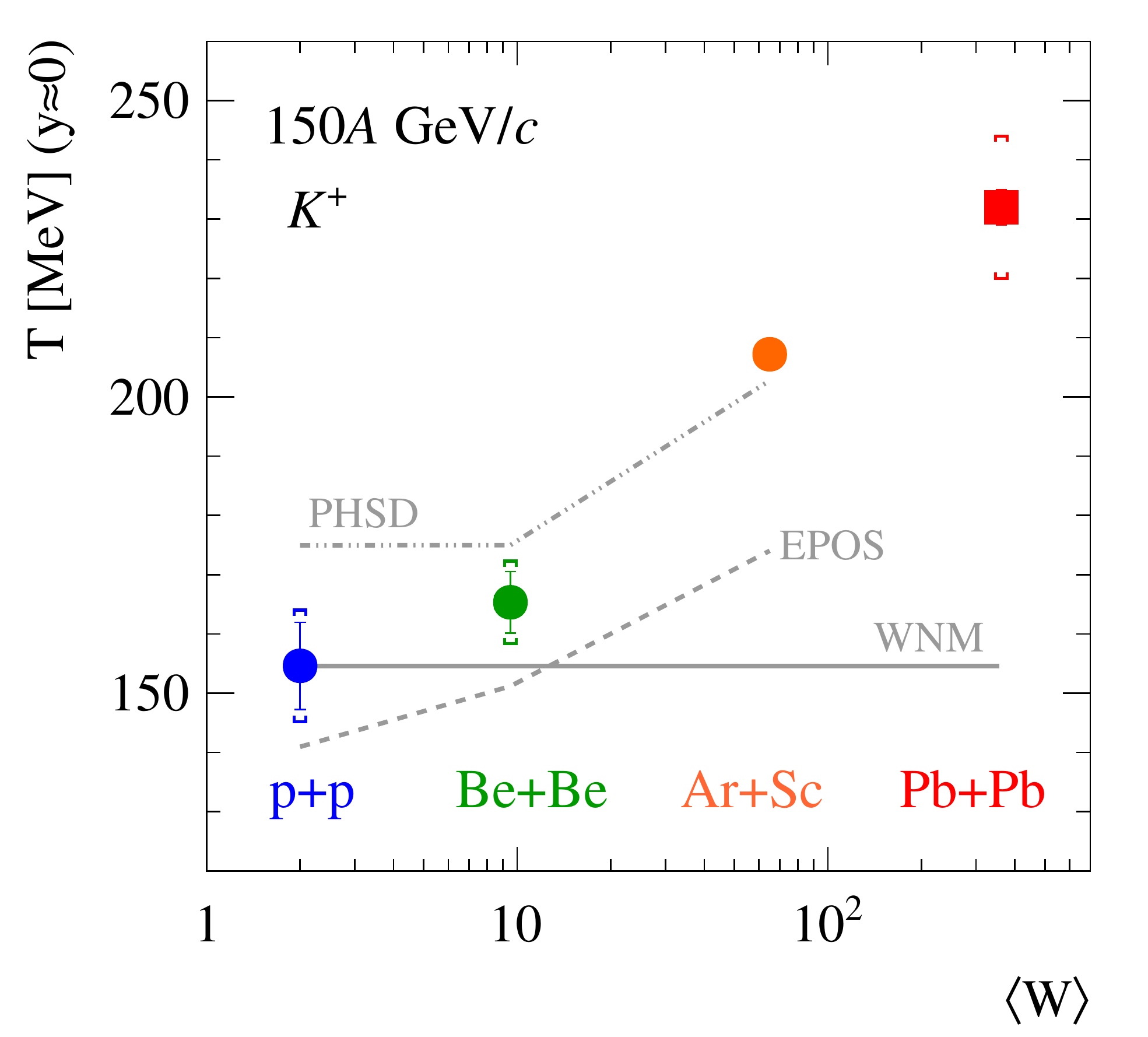}\\[-2mm]
}
\caption{The system size dependence of the $K^+/\pi^+$ ratio at midrapidity measured in collisions at 150$A$ GeV/$c$ compared with dynamical (left) and statistical (middle) models. Right: Measurement of the inverse slope parameter $T$, fitted to $K^+$ transverse momentum spectra at the same collision energy. }
\label{fig:syssize_dep}
\end{figure}

The Statistical Model of the Early Stage (SMES) \cite{2} predicts a first order phase transition from the hadronic matter to the QGP phase between top AGS and top SPS energies. In the transition region, constant temperature and pressure in the mixed phase and an increase of the number of internal degrees of freedom are expected. The most prominent predicted signature of the deconfinement transition is a rapid, non-monotonic change of the $K^+/\pi^+$ ratio as a function of collision energy (\textit{“the horn”}). Such signature was observed in central Pb+Pb collisions by the NA49 experiment, both at mid-rapidity, as well as in mean multiplicities. The NA61/SHINE extends the set of experimental data by additional measurements of lower mass systems: $p$+$p$, Be+Be and Ar+Sc.
\begin{figure}[h]
\centering
\includegraphics[width=0.33\textwidth]{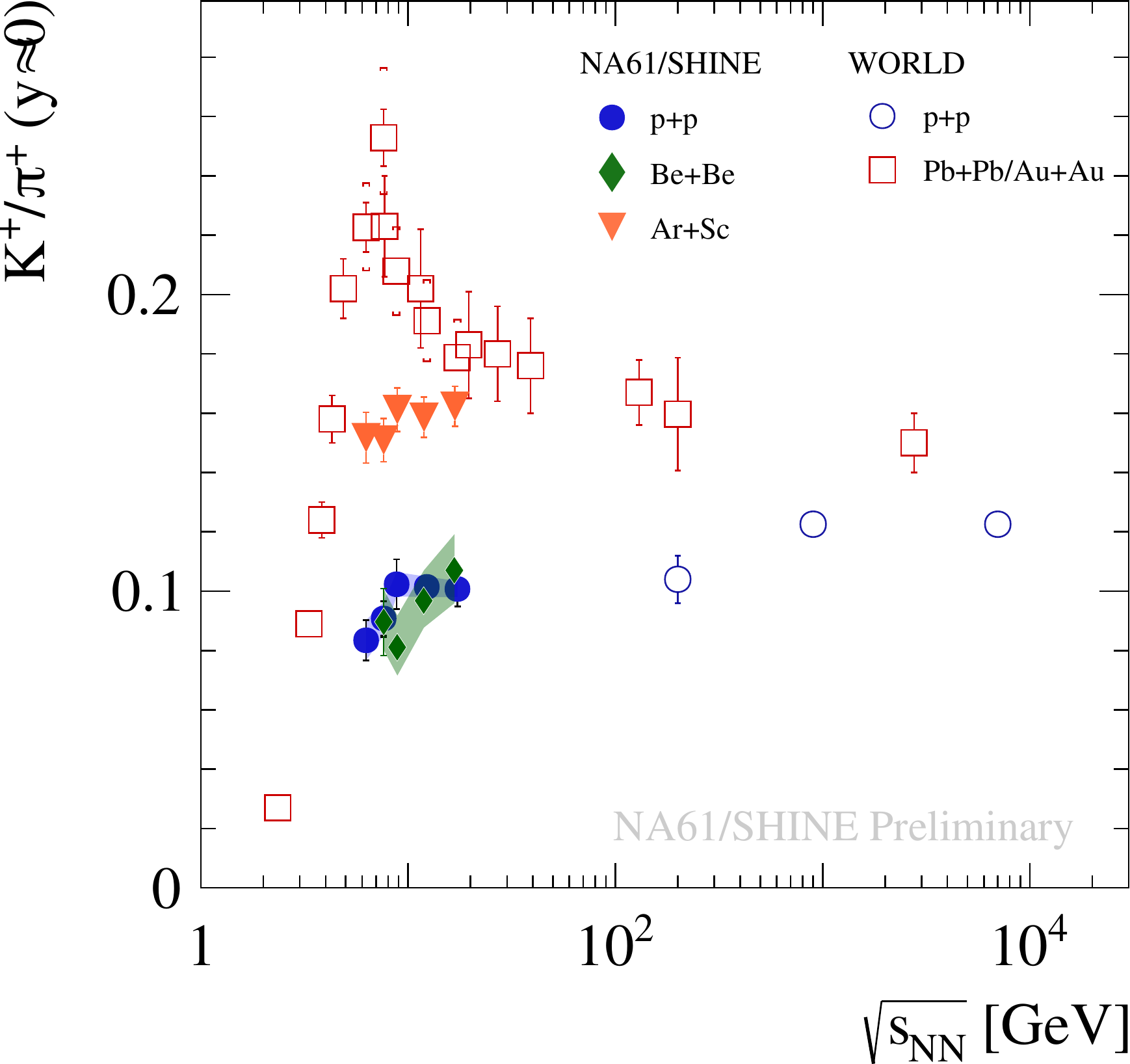}
~~~~~
\includegraphics[width=0.33\textwidth]{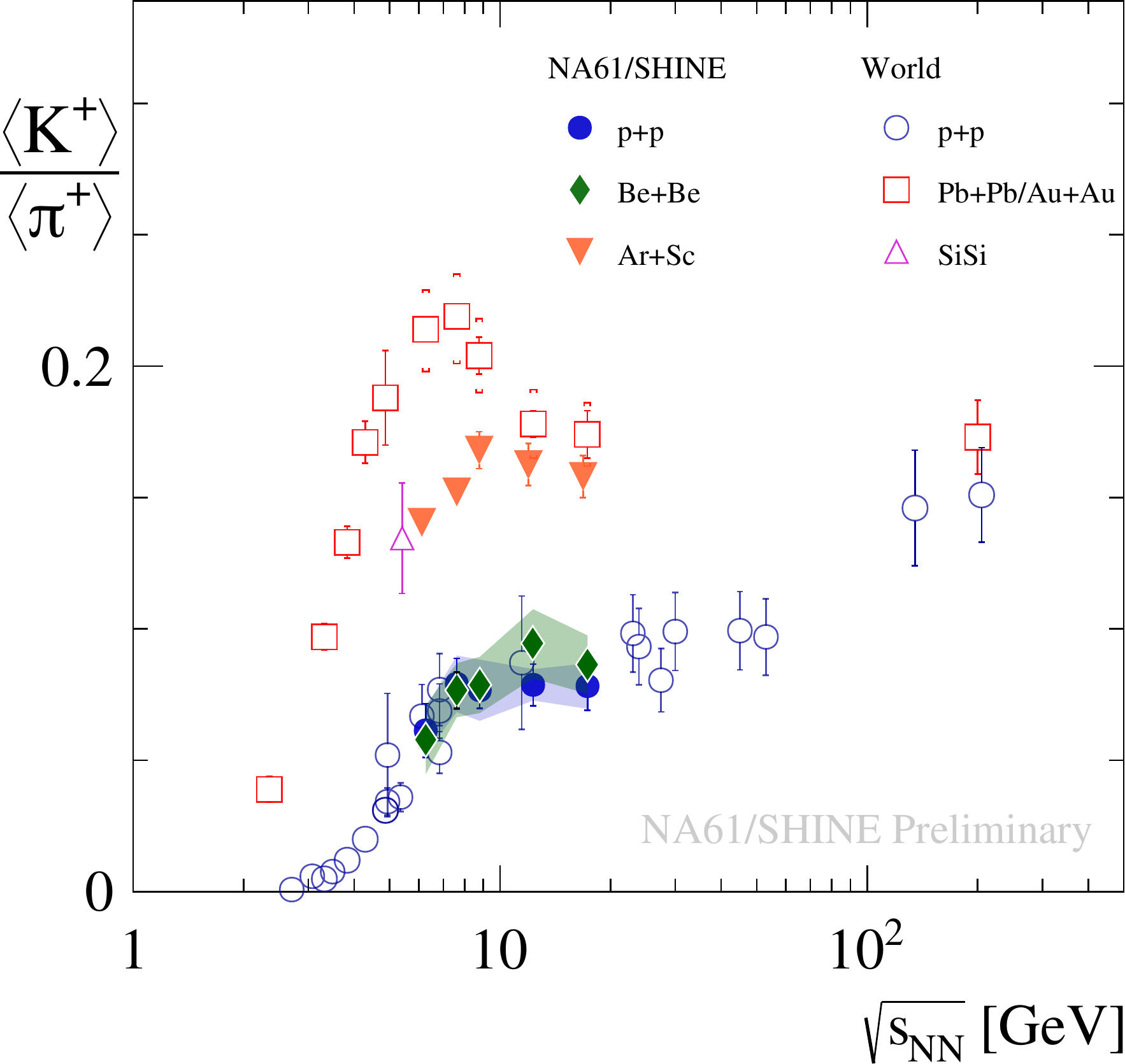}\\[-2mm]
\caption{The $K^+/\pi^+$ charged meson ratio in midrapidity (left) and for total yields (right). The plots show data on $p$+$p$, Be+Be and Ar+Sc collisions measured by the NA61/SHINE (full symbols) compared with available World data (open symbols).}
\label{fig:horns}
\end{figure}
The Fig. \ref{fig:horns} shows the compilation of results obtained by both Collaborations, as well as other available World data. A plateau-like structure is visible in $p$+$p$ interactions at SPS energy range and similar values are observed for Be+Be collisions---both at mid-rapidity as well as in the total yields. For all five analysed energies of Ar+Sc collisions, the ratio $K^+/\pi^+$ is significantly higher than in collisions of small systems ($p$+$p$, Be+Be), although with clear qualitative difference with respect to Pb+Pb data---no horn structure is visible. Figure \ref{fig:syssize_dep} shows a detailed look into the system size dependence of $K^+/\pi^+$ ratio at midrapidity and inverse slope parameter $T$ fitted to the transverse momentum spectra of $K^+$ mesons. Both shown quantities display a similar, threshold-like behaviour, which cannot be reproduced by any of considered models. Ar+Sc is the lightest of measured systems at which a significant enhancement of the $K^+/\pi^+$ ratio and increase in the $T$ parameter is observed.
  
The observed rapid change of hadron production properties that starts when moving from Be+Be to Ar+Sc collisions at top SPS energies (``the jump'') hints a beginning of the creation of large clusters of strongly interacting matter---\textit{the onset of fireball}. The similarities of $p$+$p$ and Be+Be systems suggest that interactions of these systems could happen in clusters of binary collisions of nucleons, in which produced hadrons originate from the formation, evolution and fragmentation of strings \cite{strings}. On the other hand, in case of larger systems (Pb+Pb, but also Ar+Sc), the density of strings may become sufficiently high, that the strings are close enough to interact and change their properties, forming a collectively evolving QGP fireball \cite{shuryak}.
The NA61/SHINE's studies of system size dependence of hadron production properties allows for the first time to sketch the diagrams of high energy nuclear collisions, displaying the domains in which different hadron production mechanisms dominate \cite{oof1, oof2}.

\section{New results on strange mesons produced in $p$+$p$ interactions}

\begin{figure}[h]
\includegraphics[width=0.52\linewidth]{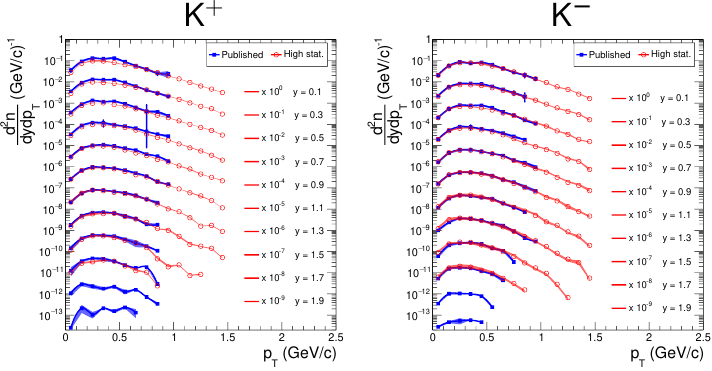}~
\raisebox{1.6cm}{
\begin{minipage}{0.4\linewidth}
\centering
\small \hspace*{-1cm}K$_S^0$ {\footnotesize(158 GeV/$c$)}\\
\includegraphics[width=\linewidth]{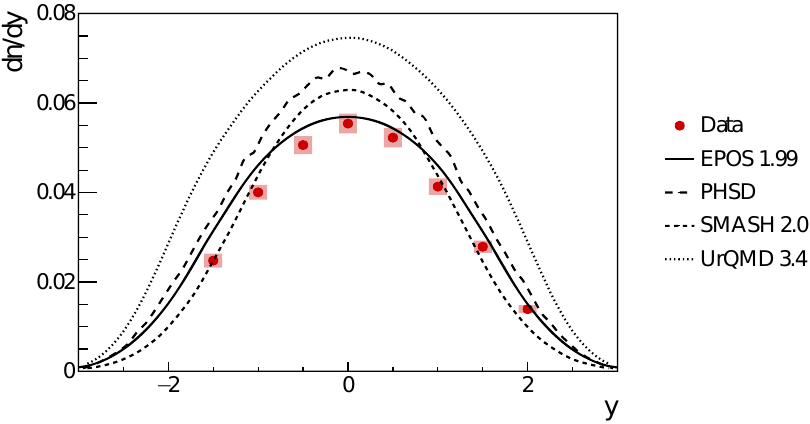}\\
\end{minipage}}
\caption{New, high-statistics NA61/SHINE results on $K$ mesons produced in $p$+$p$ inelastic interactions at 158 GeV/$c$: the two first panels show the comparison of $K^+$ and $K^-$ $p_T$ spectra (preliminary) compared to the previously published data and the plot on the right displays the rapidity distribution of the $K_S^0$ meson \cite{marjan}. }
\label{fig:kpm0}
\end{figure}

The studies of $K$ meson production in $p$+$p$ collisions is interesting not only as a reference for A+A systems, but also for understanding the strangeness production in elementary interactions. The NA61/SHINE has collected data on almost 30 million $p$+$p$ collisions, allowing for measurements of unprecedented precision at SPS energy range. The first results on $K^0_S$ mesons produced in $p$+$p$ collisions at 158 GeV/$c$ \cite{marjan} are also the first step of the energy scan of $K^0_S$ production in $p$+$p$ interactions, which will provide a baseline for the future analysis in collisions of heavier systems. Additionally, thanks to high statistics, large acceptance and good resolution, the new results on $K^0_S$, $K^+$ and $K^-$ have significantly higher precision than previously published data at the SPS energies and will become an important discrimination and tuning tool for the models of high energy collisions. The measured spectra of discussed mesons are displayed in Fig. \ref{fig:kpm0}.

\begin{figure}[h]
\raisebox{2.15cm}{
\begin{minipage}{0.33\linewidth}
\centering
~~~$K^{*}(892)^0$\\
{\small(40--158 GeV/$c$)}\\[0.1cm]
\includegraphics[width=\textwidth]{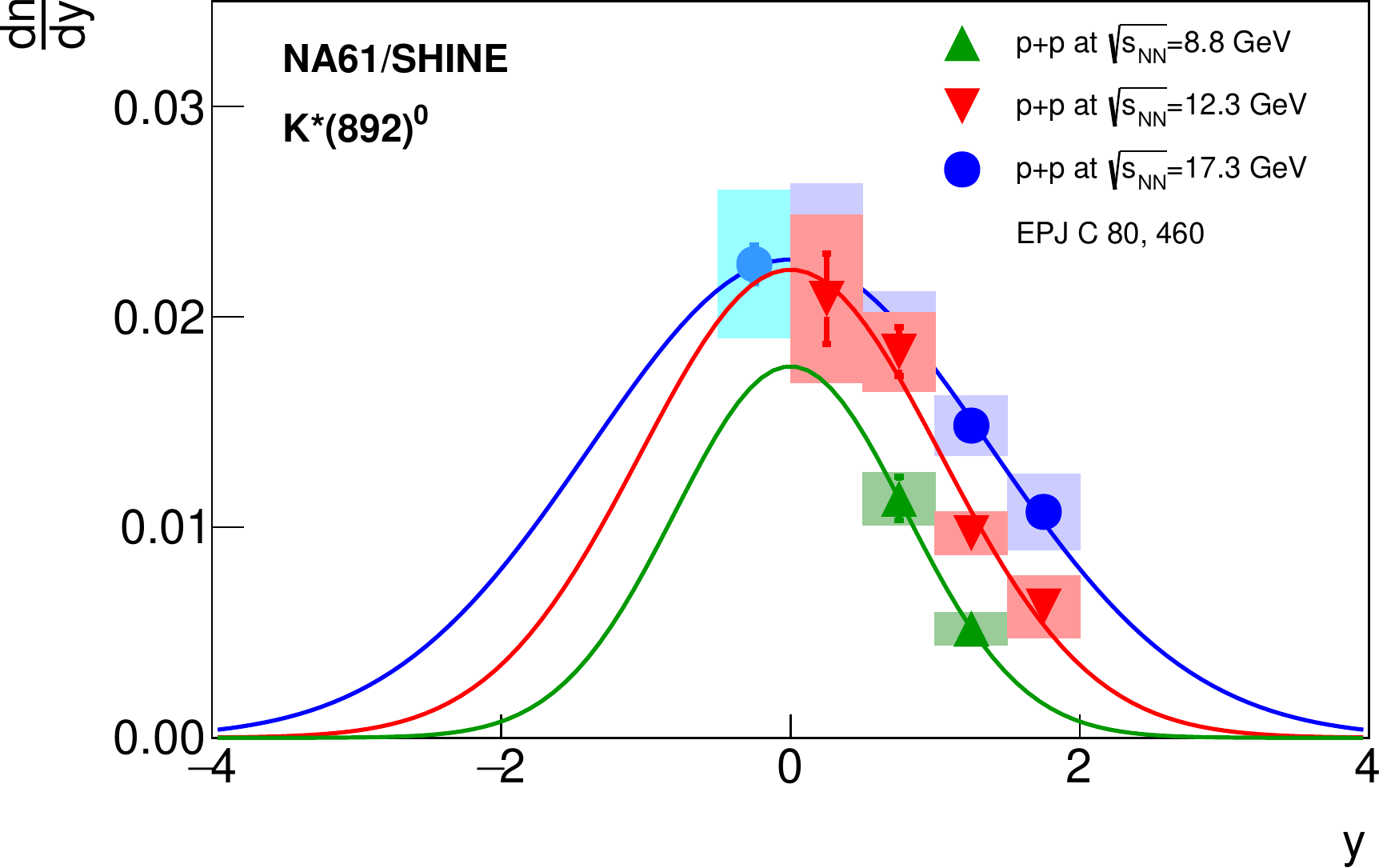} \\
\end{minipage}}~
\includegraphics[width=0.33\textwidth]{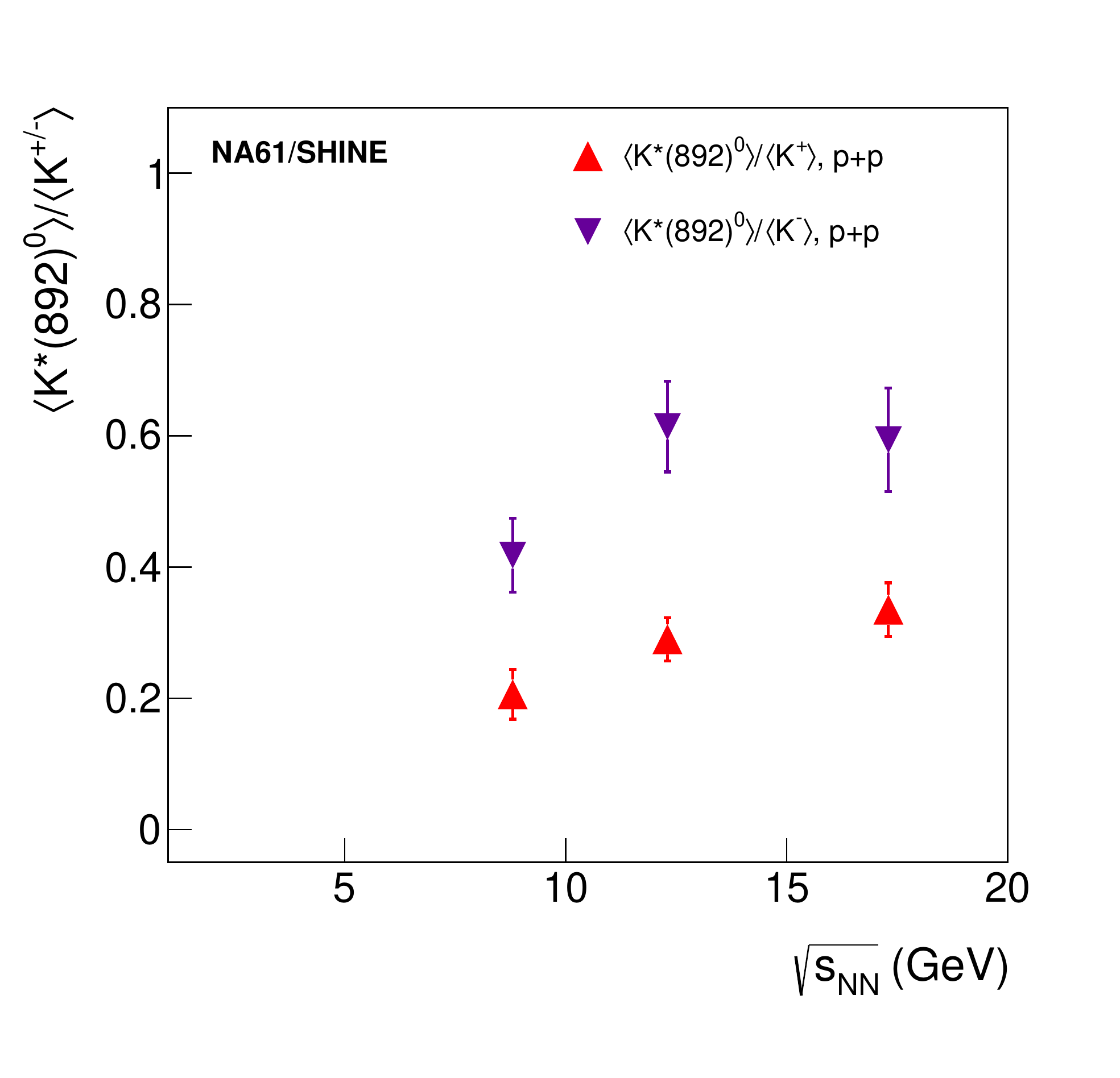}~
\includegraphics[width=0.33\textwidth]{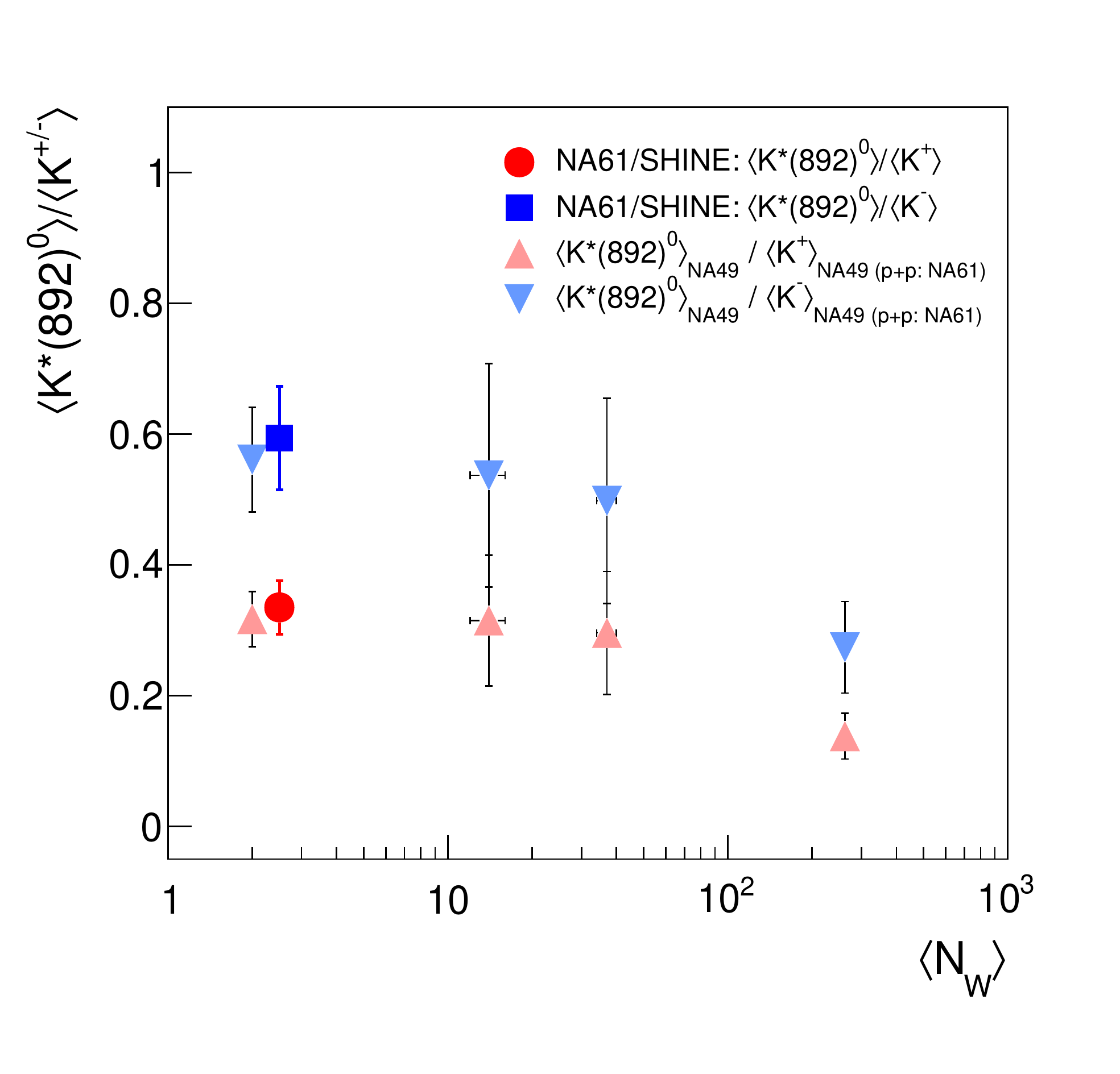}\\[-8mm]
\caption{Left: Results on rapidity spectra of $K^{*}(892)^0$ mesons produced in inelastic $p$+$p$ collisions at 40, 80, and 158 GeV/$c$. Middle: Ratio of mean multiplicity of $\langle K^{*}(892)^0 \rangle$ meson to the mean multiplicity of charged kaons $\langle K^\pm \rangle$. Right: Comparison to Hadron Gas Model predictions \cite{9}}
\label{fig:kstar}
\end{figure}

Strange hadrons may also be produced in the form of short-lived excited states. The study of such resonances is a unique insight into understanding of the time evolution of the fireball created at high energy collisions.
The transverse mass and rapidity spectra, as well as total yields of $K^{*}(892)^0$ mesons measured by the NA61/SHINE \cite{kstar1,kstar2} are displayed in Fig. \ref{fig:kstar}.
The measurement
%of $K^{*}(892)^0$ meson production
may help to distinguish between two possible scenarios for the fireball freeze-out: the sudden and the gradual one \cite{timeevo1}. Moreover, the ratio of $K^{*}(892)^0$ to the charged kaon yield (Fig. \ref{fig:kstar}) may allow to determine the time between chemical and kinetic freeze-outs \cite{timeevo1,timeevo2}. Together with future NA61/SHINE measurements in Be+Be, Ar+Sc, and Xe+La collisions, it will allow estimating the time between freeze-outs for these systems at three SPS energies. The spectra are also important inputs for Blast-Wave parametrization and Hadron Resonance Gas models, broader discussed in Ref. \cite{kstar2} and \cite{9,10}, respectively. The described high precision results on $K^{*}(892)^0$ production will also provide an important reference for tuning Monte Carlo string-hadronic models.

\section{Multi-strange hyperon production in $p$+$p$ interactions}
\begin{figure}[h]
~\\[-0.7cm]
{\centering
\includegraphics[width=0.34\linewidth]{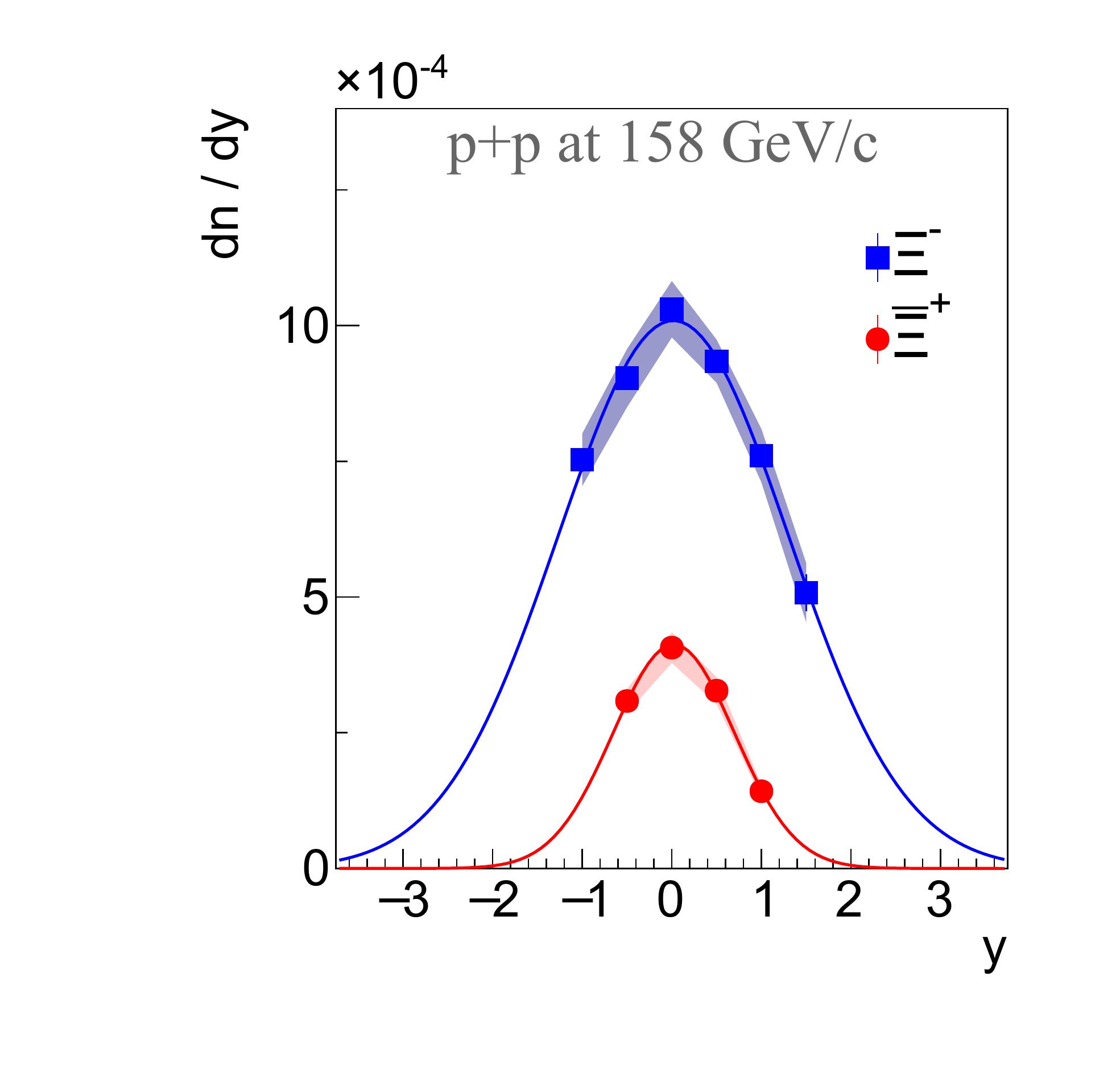}~
\raisebox{2.5mm}{
\includegraphics[width=0.36\linewidth]{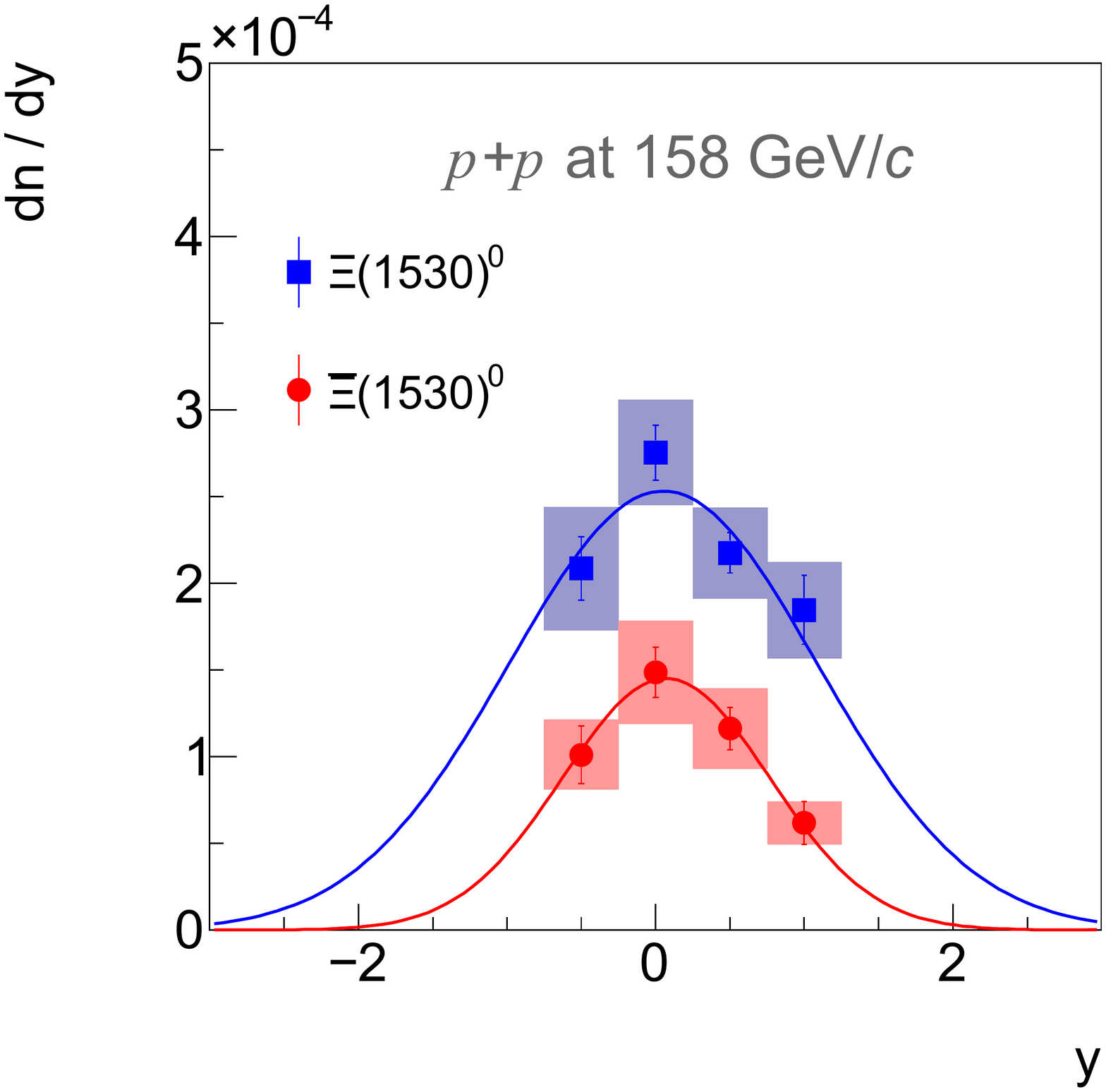}}~
\includegraphics[width=0.26\linewidth]{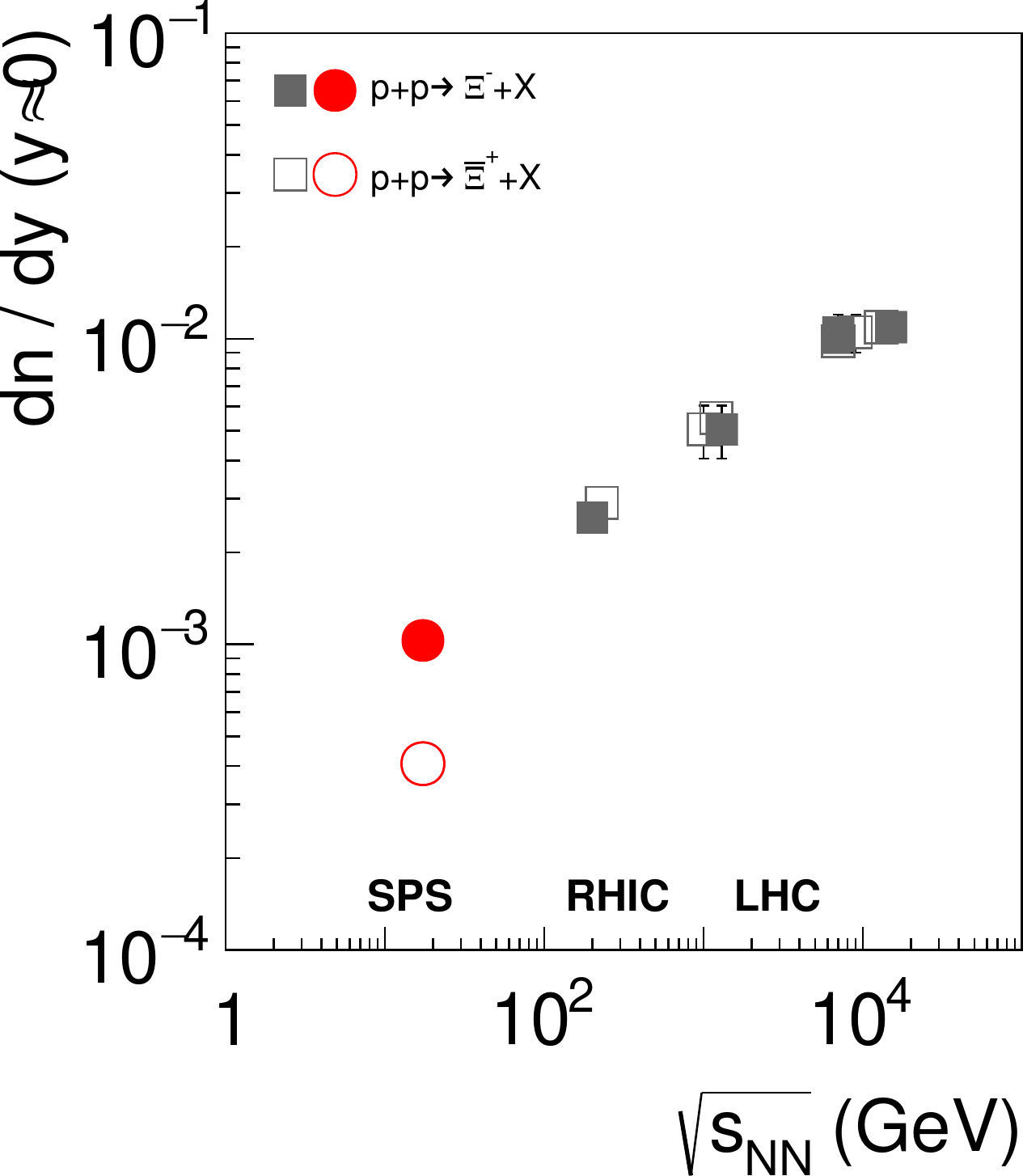}\\[-0.1cm]}
\caption{$\Xi$ hyperon production in inelastic $p$+$p$ collisions at 158 GeV/$c$: d$n/$d$y$ distributions of $\Xi^-$ and $\bar{\Xi}^+$ (left), $\Xi(1530)^0$ and $\bar{\Xi}(1530)^0$ (middle) and collision energy dependence $\Xi^-$ and $\bar{\Xi}^+$ yields at midrapidity, including new results from SPS.}
\label{fig:xipm}
\end{figure}

The high statistics sample of $p$+$p$ collisions at 158 GeV/\textit{c} beam momentum, obtained by the NA61/SHINE, allowed the more challenging measurement of $\Xi^-$, $\bar{\Xi}^+$, $\Xi(1530)^0$ and $\bar{\Xi}(1530)^0$ hyperon production \cite{xi}---the only results at the SPS energy.
Multi-strange hyperon candidates are identified by their decay typologies. The $\Xi(1530)^0$ is produced in the primary interaction and decays strongly into $\Xi^-$ and $\pi^+$. Then the $\Xi^-$ travels certain distance, after which it decays into a $\Lambda$ and a $\pi^-$. Subsequently, the $\Lambda$ decays into a proton and a $\pi^-$.
The results of described procedure are the two-dimensional spectra in $y$ and $p_T$, which are consequently corrected for detector geometrical acceptance, reconstruction efficiency, secondary interactions and finally also branching ratios to unmeasured channels. The $p_T$-integrated rapidity distribution of $\Xi^-$, $\bar{\Xi}^+$, $\Xi(1530)^0$ and $\bar{\Xi}(1530)^0$ is shown in Fig. \ref{fig:xipm}. The Figure also compiles the data on charged $\Xi$ yields at midrapidity from various experiments, showing the collision energy dependence. The figure shows a strong suppression of antibaryon yield at SPS, with $\langle\bar{\Xi}^+\rangle/\langle\Xi^-\rangle=0.395\pm0.012\pm0.030$, which disappears already at RHIC.

\begin{figure}[h]
{\centering
\includegraphics[width=0.34\linewidth]{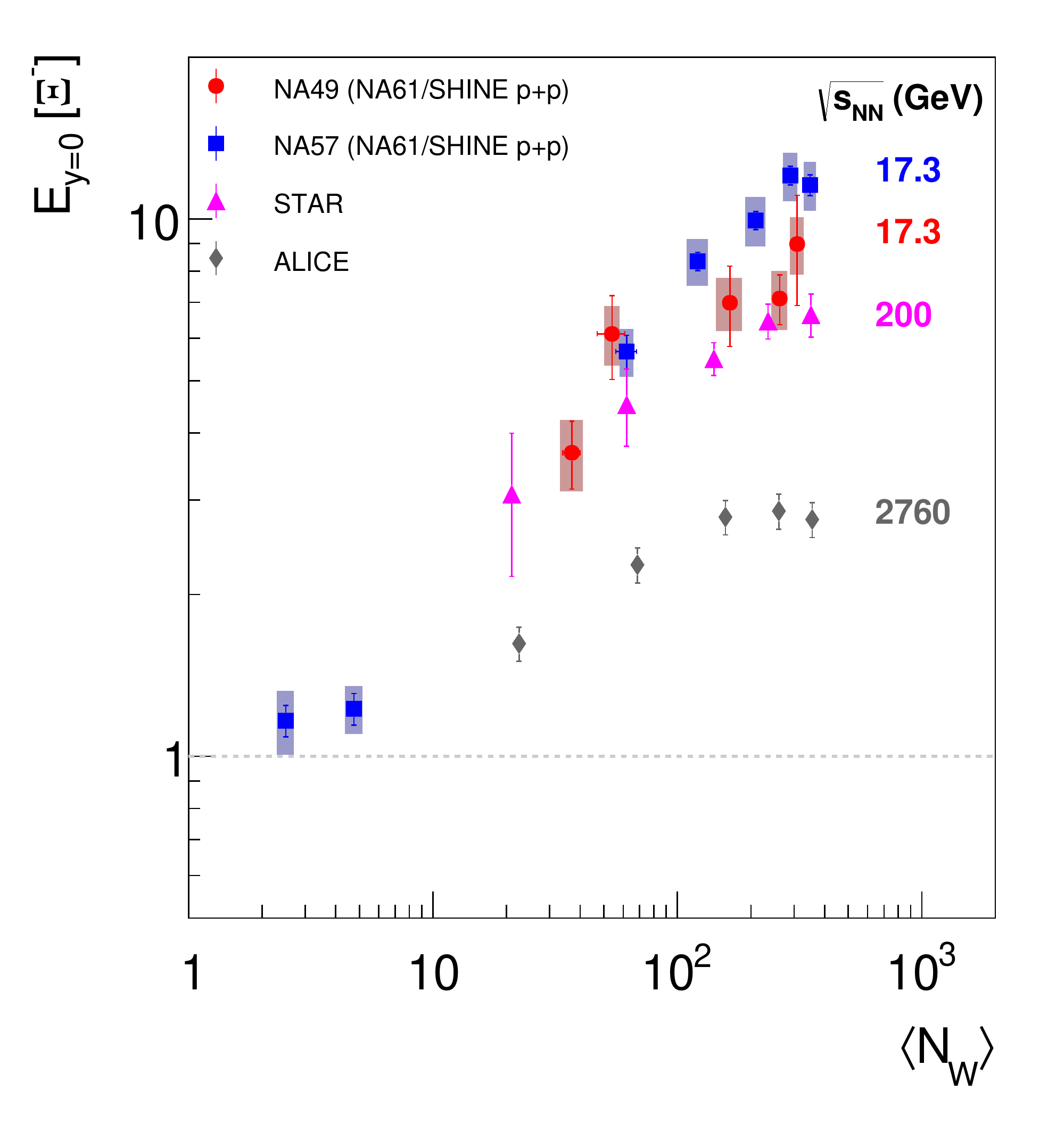}~
\includegraphics[width=0.34\linewidth]{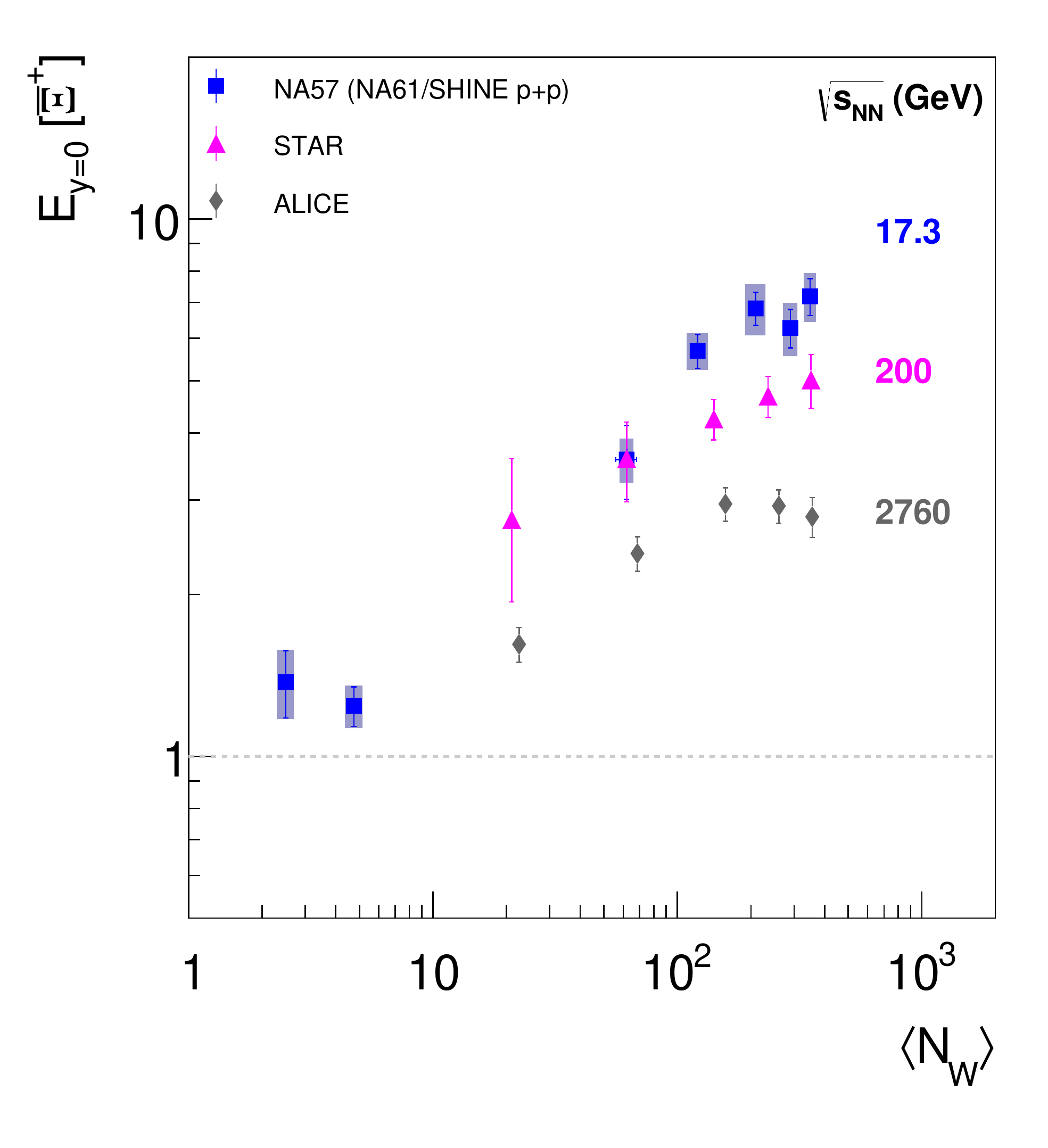}\\[-5mm]}
\caption{Strangeness enhancement factors determined using $\Xi^-$ and $\bar{\Xi}^+$ baryon production measured by the NA61/SHINE as the baseline.}
\label{fig:senh}
\end{figure}

The $\Xi$ mean multiplicities measured by NA61/SHINE in inelastic $p$+$p$ interactions are used to calculate the enhancement factors of $\Xi$ hyperons observed in centrality selected Pb+Pb, in semicentral C+C, and in Si+Si collisions as measured by NA49 \cite{na49xi} at the CERN SPS. The results for midrapidity densities are shown in Fig. \ref{fig:senh} as a function of $\langle N_W \rangle$. The enhancement
factor increases approximately linearly from 3.5 in C+C to 9 in central Pb+Pb collisions.

%\section{Summary}
%The new measurements from the NA61/SHINE Collaboration on strange mesons, baryons and hadronic resonances produced in collisions of $p$+$p$, Be+Be and Ar+Sc were presented. The most prominent behaviour observed in the analyses of collisions of intermediate size systems is that there is no horn structure visible in results on $K^+/\pi^+$ either from Be+Be, nor Ar+Sc reactions. However, an unexpected system size dependence is observed, as Be+Be values of $K^+/\pi^+$ ratio is very close to the ones from $p$+$p$, while the ratio is significantly higher in case of Ar+Sc.

~\\[-0.2cm]
{\scriptsize
\textit{Acknowledgments:} This work was supported by the Polish Minister of Education and Science (contract No. 2021/WK/10) and partially also by the Polish National Science Centre grant 2015/18/M/ST2/00125.\\[-12mm]
}


\begin{thebibliography}{99}
\scriptsize \vspace*{-5mm}
\bibitem{1}
NA61/SHINE Collaboration, JINST 9, P06005 (2014)

\bibitem{2} M. Gazdzicki, M.I. Gorenstein, Acta Phys. Pol. B 30, 2705 (1999)

\bibitem{3} NA61/SHINE, Eur. Phys. J. C 77, 671 (2017)

\bibitem{4} P. Podlaski, SQM 2019, Bari, Italy, {\tiny\url{indico.cern.ch/event/755366/contributions/3357938}}

\bibitem{5} S. Pulawski, SQM 2019, Bari, Italy, {\tiny\url{indico.cern.ch/event/755366/contributions/3427002/}}

\bibitem{6} NA61/SHINE, Tech. Rep. CERN-SPSC-2018-008,
SPSC-P-330-ADD-10

\bibitem{bebe1}
NA61/SHINE, Eur.Phys.J.C 80 (2020) 10, 961, Eur.Phys.J.C 81 (2021) 2, 144 (erratum)

\bibitem{bebe2}
NA61/SHINE, Eur.Phys.J.C 81 (2021) 1, 73

\bibitem{arsc1}
NA61/SHINE, Eur.Phys.J.C 81 (2021) 5, 397

\bibitem{strings}
K. Werner, Phys. Rept., vol. 232, pp. 87–299, 1993.

\bibitem{shuryak}
T. Kalaydzhyan, E. Shuryak, Phys. Rev. C, vol. 90, no. 1, p. 014901, 2014;\\
T. Kalaydzhyan and E. Shuryak, Phys. Rev. D, vol. 90, no. 2, p. 025031, 2014;\\
I. Iatrakis, A. Ramamurti, and E. Shuryak, Phys. Rev. D, vol. 92, no. 1, p. 014011, 2015

\bibitem{oof1}
M. P. Lewicki, L. Turko, {\tiny\href{https://arxiv.org/abs/2002.00631}{	arXiv:2002.00631}}

\bibitem{oof2}
E. Andronov, M. Kuich, M. Gaździcki, {\tiny\href{https://arxiv.org/abs/2205.06726}{arXiv:2205.06726}}

\bibitem{kstar1}
NA61/SHINE, Eur.Phys.J.C 80 (2020) 5, 460

\bibitem{kstar2}
NA61/SHINE, Eur.Phys.J.C 82 (2022) 4, 322

\bibitem{marjan}
NA61/SHINE, Eur.Phys.J.C 82 (2022), 96

\bibitem{9}
V. V. Begun et al., Phys. Rev. C 98,
no. 5, 054909 (2018)

\bibitem{10}
J. Phys. G: Nucl. Part. Phys. 48 085004 (2021)

\bibitem{timeevo1}
C. Markert, G. Torrieri and J. Rafelski, AIP Conf. Proc. 631, 533 (2002)

\bibitem{timeevo2}
C. Blume, Acta Phys. Polon. B 43, 577 (2012)

\bibitem{xi}
NA61/SHINE, Eur.Phys.J.C 81 (2021) 10, 911

\bibitem{na49xi}
NA49, Phys. Rev. C 80, 034906 (2009), 0906.0469

\end{thebibliography}
\end{document}